\documentclass{sig-alternate}
\begin{document}


\CopyrightYear{2017} 
\setcopyright{licensedothergov}
\conferenceinfo{SAC 2017,}{April 03 - 07, 2017, Marrakech, Morocco}
\isbn{978-1-4503-4486-9/17/04}\acmPrice{\$15.00}
\doi{http://dx.doi.org/10.1145/3019612.3019795}





%

\title{Security and Privacy Preserving Data Aggregation in Cloud Computing}

\numberofauthors{1}
\author{
%
%
\alignauthor Leandro Ventura Silva, Rodolfo Marinho, Jose Luis Vivas, Andrey Brito\\
       \affaddr{Universidade Federal de Campina Grande}\\
       \email{\{leandrojose, rodolfoams, jlvivas, andrey\}@lsd.ufcg.edu.br}
}


\maketitle
\begin{abstract}
Smart metering is an essential feature of smart grids, allowing residential customers to monitor and reduce electricity costs. Devices called smart meters allows residential customers to monitor and reduce electricity costs, promoting energy saving, demand management, and energy efficiency.  However, monitoring a households' energy consumption through smart meters poses serious privacy threats, and have thus become a major privacy issue.
Hence, a significant amount of research has appeared recently with the purpose of providing methods and mechanisms to reconcile smart metering technologies and privacy requirements. However, most current approaches fall short in meeting one of several of the requirements for privacy preserving smart metering systems. In this paper we show how Intel SGX technology can be used to provide a simple and general solution for the smart metering privacy problem that meets all these requirements in a satisfactory way.  Moreover, we present also an implementation of the proposed architecture as well as a series of experiments that have been carried out in order to assess how the proposed solution performs in comparison to a second implementation of the architecture that completely disregards privacy issues.
 
\end{abstract}

\ccsdesc[500]{Security and privacy~Privacy-preserving protocols}
\ccsdesc[500]{Security and privacy~Hardware security implementation}
\ccsdesc[500]{Security and privacy~Security protocols}
\ccsdesc[500]{Security and privacy~Data anonymization and sanitization}
\ccsdesc[500]{Security and privacy~Information accountability and usage control}
\ccsdesc[500]{Security and privacy~Privacy protections}

%
%

%
%
\printccsdesc


\keywords{smart grid; smart metering; secure cloud computing; privacy}

\section{Introduction} \label{sec:introduction}

Electricity is today the most important single source of carbon dioxide emission, and hence it is considered one of the main contributors to climate change. Growing energy needs are therefore forcing governments to find more efficient and economic ways to manage the energy grid and improve load balancing.  
To this end, the current electrical power system is undergoing significant modernization with the introduction of information technologies, turning into what is commonly called ``smart grid''. 
Smart grids are basically power networks enhanced with information technology and an intelligent metering infrastructure. 
The purpose is to integrate the behavior and actions of all involved actors and components in order to ensure an economically efficient, robust, secure and reliable power system with minimal power loss. 
Among its benefits is a quicker response for changes in electricity demand, made possible by the use  of intelligent electronic devices such as load controllers, sensors, and smart meters.

Smart metering is an essential feature of smart grids, allowing residential customers to monitor and thereby reduce electricity costs.
The key component of a smart metering system is an electronic device called \textit{smart meter} (SM), which is  intended to substitute ordinary electromechanical meters that provide only energy measurement data. Unlike the latter, a smart meter can record energy consumption at determined time intervals and report it to the utility supplier for purposes of monitoring and billing. 
This is done by means of a two-way communication link between the SM and other components of the metering system. 
In theory, SMs could be capable of communicating with any other devices, including other SMs, and of executing both local and remote command signals. 


The ability to remotely read fine-grained measurements or energy consumption is the most important feature of smart metering. 
This feature would enable grid operators to perform efficient load balancing and to offer customers time-dependent tariffs. 
In this way, SMs may be used to promote energy saving, demand management, and energy efficiency. 
However, monitoring a households' energy consumption through SMs poses serious privacy threats, and have thus become a major privacy issue.
Fine-grained consumption data could reveal what citizens do within their own homes, how they spend their free time, whether they are away on holiday or at work, when they watch TV or run a washing machine, or even the use of some specific medical device. 
There is also a potential for extensive data mining by combining fine-grained metering with data from other sources. 
Notwithstanding, the benefits of smart metering are so significant that it has given rise a significant amount of research  with the purpose of providing methods and mechanisms to reconcile smart metering technologies and privacy requirements. 

The two most important smart metering privacy requirements are: (i) fine-grained individual consumption data must be accessible only to the corresponding individual; and (ii) the consumption data of a customer for a billing period (monthly or other) must be accessible to the contracted utility supplier for billing purposes. 
hence, fine-grained data metering, also called \textit{operational metering}, should not be attributable, whereas long-term consumption data, also called \textit{accountable metering}, should be attributable.

On the other hand, among the most important requirements for smart metering privacy preserving systems are: (i) scalability; (ii) fault tolerance; (iii) reduced computational and storage capabilities in smart meters; (iv) low communication overhead; and (v) rapid response. As we shall see, most current privacy preserving proposals fall short in meeting one of several of these requirements.

In the paper we show that Intel SGX technology can be used to provide a simple and general solution for the smart metering privacy problem detailed above.

\section{Intel SGX}
Intel SGX can be described as a set of new instructions and changes in memory access mechanisms added to the Intel Architecture that allows an application to instantiate a protected area in the application's address space, dubbed an \textit{enclave}~\cite{mckeen2013innovative}.
An enclave is intended to ensure confidentiality and integrity of owned data even in the presence of privileged malware. 
Inside  the  enclave,  code, data and stack  are  protected  by  hardware  enforced  access  control  policies which prevent attacks against the  enclave's content even when these originate from privileged software such as virtual machine monitors, BIOS, or operating systems. 

Another critical feature of Intel SGX is a set of instructions for remote attestation of a running enclave~\cite{anati2013innovative}. 
SGX provides means to  generate  a  hardware  based  attestation that an enclave has been successfully established on an SGX enabled platform and that a specific  piece of  software, and nothing else, has been loaded within the enclave. 
Both features, i.e., attestation and isolation, are essential to the solution we propose in this paper. 

Possible applications of Intel SGX have been discussed in \cite{hoekstra2013using}, where  examples of applications that make use of the capabilities of Intel SGX were presented, as well as an application architecture including an application split between components requiring security protection which should run within enclaves, and components that do not require protection and can therefore be executed outside enclaves.

In \cite {barbosa2016foundations} a set of tools is presented for specifying protocols that extend the guarantees provided by isolated execution environments, such as the Intel SGX, by means of its ability to perform remote attestation. In this way, it becomes possible to establish secure key exchange protocols between a remote participant and an isolated execution environment. These protocols are defined by the combination of a passively secure key exchange protocol and an arbitrary attestation protocol.

\section{Privacy-preserving solutions for operational metering}
\label{sec:current_solutions}

Basically, privacy preserving proposals for smart metering have adopted one of the two following approaches: anonymization and data aggregation. Next, we present each one of them.

\subsection{Anonymization approaches}

Anonymization approaches have been based on trusted third parties (TTP), zero knowledge proofs, and group keys/IDs.
 
\paragraph*{Trusted third party solutions} TTP solutions include the following proposals: (i) using a trusted neighborhood gateway to which SMs send their data at each time slot, and which thereafter forwards to the utility supplier only the metering data without the corresponding identity information \cite{Molina-Markham:2010:PMS:1878431.1878446}; (ii) using a Public Key Infrastructure (PKI) for the identification of smart meters to the TTP, in which a SM may send to the TTP its private data encrypted with both the public key of the supplier and a pseudonym private key shared with the TTP, which may thus authenticate the data but not read it before sending it to the utility supplier encrypted with the latter's public key, without any identification details \cite{6165271}; (iii) anonymization of operational metering data through pseudonymous certified by a TTP escrow service \cite{citeulike:13267453}; and (iv) letting a trusted proxy, either the gateway or a TTP, play the role of an anonymizer hiding the static IP address of SMs \cite{DBLP:journals/tsg/GongCGF16}.

Apart from vulnerabilities such as timing attacks, the main drawback of these anonymization approaches is related to the trust model issues: in each case a TTP must be trusted. We need solutions that minimize the required trusted base.

\paragraph*{Zero Knowledge proofs}

Instead of relying on TTPs, other schemes have been proposed that are based on the SM sending metering data in plaintext to authorized data recipients using zero-knowledge methods to anonymously prove that it is a legitimate SM, but without revealing any other information, such as the identity of the SM \cite{journals/iacr/QuSLS15,secrypt11}. The main drawbacks of these approaches are the high computational costs and communication overheads induced by zero-knowledge proof methods and the use of anonymous credentials.

\paragraph*{Group keys and IDs}
 
An anonymization approach was proposed using a virtual ring architecture in which SMs are organized, based on their location, into groups, each group with a group ID and a pair of public/private ring keys used by each SM to sign its operational metering data before sending it to the utility supplier, which would thus be able to verify the authenticity of the SM but not its ID \cite{6697806}. A limitation of this approach is lack of scalability, since the message signature size increases linearly with the increasing group size. 

Another proposed scheme is based on group keys, HMAC signatures, PKI, trusted substations, and the inclusion of a trusted entity called \textit{control} center (the power generator system)  \cite{conf/smartgridcomm/ChimYHL11}. 
The control center generates a system key and distributes it to the SMs of a region and its corresponding substation, which may then request more electricity by sending to the control center a fresh pseudonym in the shape of a random number concatenated with their real ID encrypted with the control center's public key. 
The main drawback of this  scheme is that the SMs protect their identities from the substations, but not from the control center, and hence from the utility itself. 
This was later improved in order to guarantee that the real identities of the SMs remain hidden also to the control center, but in this case introducing high a communication overhead \cite{DBLP:conf/globec/CheungCYLH11}, a drawback which was eliminated later \cite{journals/cm/ChimYHL12}. An improvement was also proposed by replacing the blind signature with a ring signature \cite{Yu2014PrivacyPreservingPR}, reducing computational costs and communication overheads, but at the price of enabling the control center to link electricity requests via a static pseudonym, and ultimately to a real ID.

Another proposal \cite{conf/trustcom/StegelmannK12} makes use of a non-trusted transportation layer $k$-anonymity service (the Gateway Operator) using pseudonyms, each one used by at least $k$ SMs for sending encrypted metering data to the utility provider and signed for the Gateway Operator, who verifies the signature, removes it, and sends the encrypted message to the utility provider. However, over a long period of time it has been shown that long term aggregation of partial information can break the anonymity of the users. 

Finally, a proposal relying on a pseudonymous smart metering protocol without a TTP was put forward in which each SM generates its own blinded  pseudonym, gets a signature of it from the service supplier, and thereafter unblinds the received signature \cite{conf/trustcom/FinsterB13may}. In this way, a SM gets a pseudonym that is signed by the utility supplier but that the latter cannot associate with a specific SM. A drawback here is that new SMs joining the system could have their new pseudonyms easily linked to themselves.

\paragraph*{Conclusions on anonymization approaches} 

As we may see from the presented proposals, there is usually a trade-off between the extent of the trusted base and the complexity of the solution, either in terms of computation or communication overheads.  Apart from the higher computational and communication overheads associated with pseudomization schemes, those schemes are vulnerable to de-anomization attacks based on statistical analysis, correlation of account and operational data, and side-channel attacks.

\subsection{Data aggregation approaches}

For operational purposes the value of the aggregate consumption metering data of all the customers in a given region would be enough to guarantee the benefits of smart metering. It is assumed that it would be difficult for the utility supplier to decompose the individual users' metering data from the aggregated data, thus ensuring their privacy.

Three types of data aggregation approaches may be distinguished, based on (i) trusted third parties, (ii) data perturbation, and (iii) cryptography.

\paragraph*{Trusted third parties (TTPs)} 
In these approaches, TTPs may collect the raw metering data of all the users, aggregate these data and  deliver only the aggregated metering data to the utility supplier or other authorized data recipients, ensuring users' privacy against the utility supplier, but not the TTP, which receives all the metering data of the individual users \cite{bohli10privacy}. The use of pseudonyms known only to the supplier has been suggested to prevent it, as shown in the previous section. However, the use of pseudonyms leads to the drawbacks of the anonymization approaches discussed above. Another scheme uses the Diffie-Hellman key exchange protocol for establishing a secret key shared between an SM and network Gateways \cite{journals/tsg/FoudaFKLS11}.

The obvious flaw in these types of schemes is that solutions must rely on a TTP.  

\paragraph*{Data perturbation} 
The approach here is that each SM adds random noise to its metering data so that the aggregator is not be able to find out the original metering data but may still calculate the aggregated value with a small or negligible expected error \cite{bohli10privacy}. Several types of random error distributions have been proposed in the extensive literature on the subject. The major drawback of these approaches is the difference between the perturbed and real data and lack of fault tolerance, since the failure of a single SM to deliver its data may prevent obtaining a good estimation of the real value.

\paragraph*{Cryptography} 

The use of cryptographic primitives has been proposed to overcome the drawbacks of the previous methods. Two 
approaches are common here, one based on secret sharing schemes, and other on homomorphic encryption.

In the case of secret sharing schemes, measurements are completely hidden from the aggregator since it receives only encrypted measurements that it cannot decrypt and random shares of the total consumption \cite{conf/stm/GarciaJ10}. In this scheme, each SM splits its measurements into $k$ random shares, where $k$ is the number of SMs in a given set, including itself, keeps its own share and sends to the aggregator the remaining shares encrypted with the public keys of the corresponding SMs. The aggregator will then add the shares encrypted with one single key, using the homomorphic property of the encryption scheme, and sends the encrypted aggregated value to the corresponding SM that may then decrypt it with its secret key, add its own share, and send the result to the aggregator, which upon receiving the sums from each SM may obtain the total consumption value. The drawbacks with this scheme are mainly that it is not scalable, since it relies on secret sharing, and that the communication overhead is significant.

Other approaches use masking in such a way that when the masked inputs from all parties are summed, the masking values cancel each other out and the aggregator may thus obtain the total consumption value, maybe using simple brute force methods \cite{Kursawe:2011:PAS:2032162.2032172}. 
The communication and computation overhead of this protocol is nevertheless significant.
Other approaches are based on computing the aggregated consumption using a version of the Paillier cryptosystem \cite{Erkin2012}. This scheme requires SMs to be able to perform operations such as  Paillier encryption, hash functions, random number generation, and to communicate with each other. 
Finally, a solution relying on an additive homomorphic encryption scheme that avoids the need for secret sharing or public cryptosystems has also been presented \cite{Acs:2011:IDD:2042445.2042457}. Encryption here is simple, consisting only of a modular addition operation. The aggregator can be any SM, and the utility supplier cannot observe the actual measurements because each one is masked by a set
of random numbers that cancel each other out when they are added together. This scheme requires, however, that each smart meter shares keys with the utility supplier and exchange pseudo-random numbers among themselves. Moreover, the aggregator must be trusted.

\section{The Intel SGX solution to privacy preserving smart metering}

Models for privacy in smart metering typically include the following components: (i) \textit{consumers}, the end-users receiving the power supply, whose privacy must be protected; (ii) the \textit{Aggregator}, which receives the consumers' metered consumption from SMs, aggregates them and sends the resulting aggregated values to  the utility supplier; (iii) \textit{Smart Metering Devices} installed at the customer side of the network that periodically sense the consumed energy and send the measurements to the consumer and/or the aggregator; (iv) the \textit{Utility Supplier}, which is the company responsible for electricity distribution and transportation infrastructure, and whose operators may employ smart metering data in order optimize the provision of electricity and load balancing; and (v) the \textit{communication network} enabling two-way communication among the parties in a smart grid. 

Borrowing from the notation introduced in Bohli et al. \cite{bohli10privacy}, our model includes a set 
$S = \{SM_1,...,SM_n\}$ of smart metering devices (SMs).  We let $e_{i,j}$ denote the value of the electricity consumption of the smart meter $SM_i$ in the period $j$ for $j \in \{1,...,t\}$, where $t$ denotes the total number of time slots in one billing period. The metering values in a smart metering period can thus be described by the matrix $(e_{i,j})$. The model assumes that each value $e_{i,j}$ is unknown to the utility supplier, whereas the  aggregated values $\sum_{j=1}^t e_{i,j}$ for all $j \in \{1,...,t\}$ (the sum of electricity consumption values of an individual
customer over a given time period) and $\sum_{i=1}^{|S|} e_{i,j}$ for all $i \in \{1,...,|S|\}$ (the sum of the current electricity consumption of all customers) may be considered as public and the only values that the utility supplier needs to know.

In the trust model of our solution, the only trusted component will be the SM devices. The aggregator will include an SGX enclave component for performing aggregation which, rather than trusted, is assumed to be secure.

\begin{figure}[ht]
\centering
\includegraphics[width=.45 \textwidth]{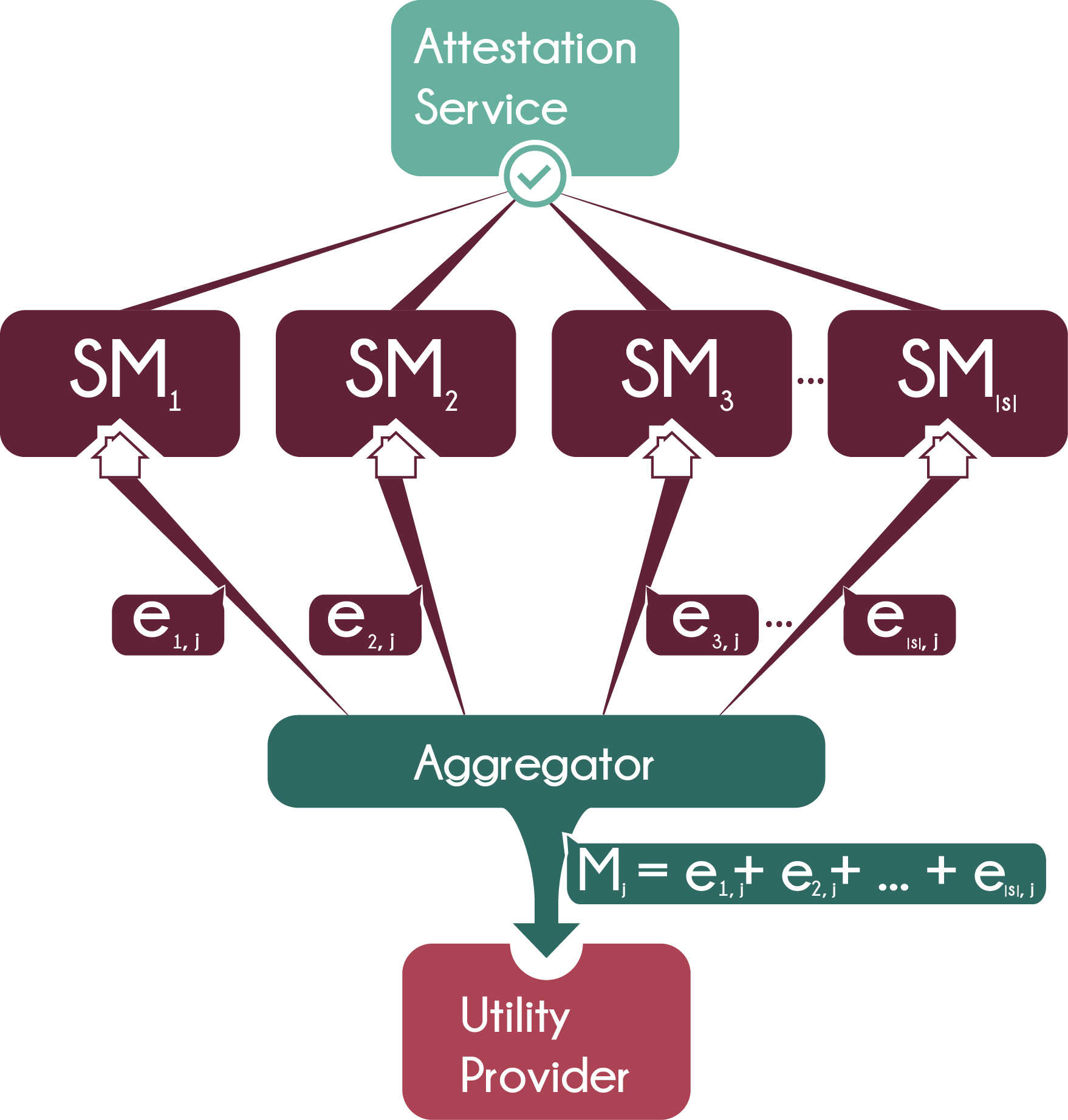}
\caption{Simplified scheme of the proposed SM architecture}
\label{fig:SM-components} 
\end{figure}

The architecture of our solution is depicted in Figure~\ref{fig:SM-components}. 
This figure shows one round in the aggregation of metering values for a given period of time $j$. Each smart meter $SM_i$ sends the current metering value along a secure communication channel that has been previously established between the aggregator and each smart meter. 
The aggregator collects these values, aggregates them, and sends the aggregated value to the utility provider. 
The aggregator will not reveal the individual values because the aggregation happens in a protected enclave that has been attested by each SM individually. 
The verification of the attestation may happen with the aid of a trusted component, the Attestation Service, which can be the Intel Attestation Service\footnote{https://software.intel.com/en-us/blogs/2016/03/09/intel-sgx-epid-provisioning-and-attestation-services} or any other trusted agency. 
The smart meters could also carry out this verification themselves. 
It is assumed that the SMs themselves are trusted by the customers, and have been previously verified. 
At the end of a billing period, the aggregator could also provide the aggregated consumption value of each smart meter to the utility provider. 
The code for this operation should also be contained within the attested enclave.

\begin{figure}[ht]
\centering
\includegraphics[width=.45 \textwidth]{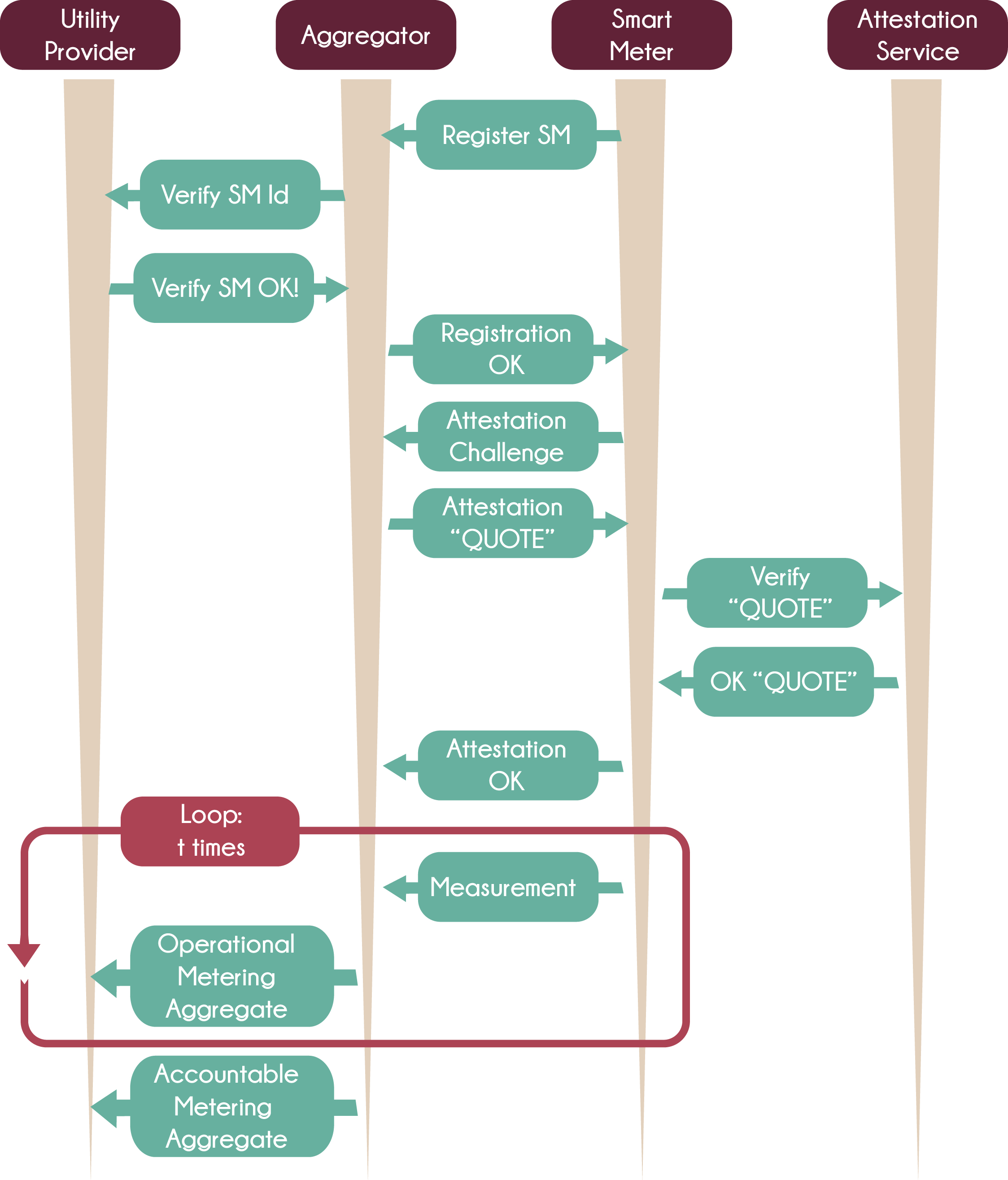}
\caption{Sequence diagram of the main scenario}
\label{fig:SeqDia} 
\end{figure}

Details about the sequence of events in the main smart metering are shown in Figure~\ref{fig:SeqDia}. 
In this scenario we include the the actors appearing in Figure~\ref{fig:SM-components}, but only a single smart metering device (SM). 
It is assumed that the SM can identify itself, maybe by means of a PKI key pair configured by the Utility Provider (UP) at deployment time. 
The sequence starts with the SM identifying itself to the Aggregator by means of some predefined identification credentials. 
The Aggregator sends the credentials to the UP for verification. 
If the action succeeds, the Aggregator reports OK to the SM, which will then proceed to attest the Aggregator to see if it's a legitimate Intel SGX platform and that  it is running an enclave with the desired code.  
After the attestation challenge has been sent, the Aggregator responds by sending back a QUOTE structure signed with an Intel Enhanced Privacy ID (EPID) key used to sign enclave attestations \cite{brickell2011enhanced}. 
The QUOTE structure contains the information required for the attestation of an enclave. 
Several other communication events may happen between the Aggregator and the SM at this juncture, e.g., actions to exchange symmetric keys in order to establish a secure channel between the SM and the Aggregator, but for the sake of simplicity these are not shown in the figure.  
If the verification of the attestation succeeds, the SM may report it to the Aggregator, upon which it can start sending periodically the fine-grained values of electric consumption along a secure channel which has been established during the attestation process. 
The Aggregator is supposed to collect the received metering values from each SM in the system, and return the aggregated value to the UP. 
At the end of a billing period it may also return the aggregated consumption value of each SM, and start over for a new period.

\subsection{Evaluation of the proposed model}

As we have seen in Section ~\ref{sec:current_solutions}, most current solutions to the smart metering privacy problem fall short in some of the most important requirements for privacy and performance. 
Our solution meets all those requirements in a simple way and with a minimum trusted base. 
We consider these requirements here one by one.

  \paragraph*{Scalability} The system is clearly scalable, since only simple  communication operations on unencrypted values are carried out by involved parties.
\paragraph*{Fault tolerance} The Aggregator does not need to wait for the consumption values of all smart meters. If one or more of the SMs fails temporarily, the aggregation may be performed with, for instance, some average value for the latest received consumption values from the corresponding SM. A threshold on the maximum fraction of failed instances at any given time slot may be defined.
\paragraph*{Limited computational and storage capabilities} Only common cryptographic operations need to be carried out by the smart metering devices, basically the ones required for establishing secure channels.  
\paragraph*{Low communication overhead} Only some few simple procedures need be performed by the smart metering devices for establishing attestation. Smart metering values are communicated at once, probably encrypted with the symmetric key exchanged during the attestation process. 
The smart meters do not need to know or communicate with each other, and no further procedures are required.
\paragraph*{Rapid response} No time consuming operations or procedures are required. 
Aggregation is done on data in the clear and only a simple summation operation is required.

For these reasons, we consider the presented solution as the most appropriate one presented so far for preserving privacy in smart metering systems. 
We assume only that Intel SGX technology is secure. 
The attacks that can be carried out against the proposed protocol are basically those that can be directed against Intel SGX (for a detailed account of possible threats and vulnerabilities in Intel SGX, see \cite{cryptoeprint:2016:086}).

Being a novel technology, the security of Intel SGX itself is still an open issue.

\section{Performance Assessment} \label{sec:performance}


For assessing the performance of the proposed architecture, two different implementations were made in C++, the first one a simple non-secure implementation without any privacy guarantees or other overhead, and the second one an implementation of the privacy preserving Intel SGX based architecture presented above. For comparison purposes,  in both implementations the same test case was used, i.e. household energy consumption aggregated by region. 

In the first implementation, all the measured data is sent by the corresponding smart meters in plain text to the aggregator, which thereafter aggregates all measurements and sends it to the utility supplier. In the second one, based on the Intel SGX technology, each smart meter, before sending its consumption measurement data, will first identify itself and thereafter attest remotely the aggregator. 

For communication between smart meters and aggregators, a bus service solution provided by Apache Kafka\footnote{http://kafka.apache.org} was used, which ensures   scalability, fault tolerance, and performance.

In order to allow communication between the C++ applications and   Apache Kafka, the library \textit{librdkafka}\footnote{https://github.com/edenhill/librdkafka} was used. Additionally, the remote attestation process is carried out by means of  a sequence of RESTful calls using the Restbed framework\footnote{https://github.com/Corvusoft/restbed}. Every attestation message is represented by a JSON\footnote{http://www.json.org/}, a data-interchange format.

The tests were conducted in a private cloud using OpenStack for the orchestration of the cloud application with Docker containers (using the OpenStack driver \textit{nova-docker}). The containers executed Ubuntu Linux 14.04 and the Intel SGX drivers\footnote{https://01.org/intel-softwareguard-eXtensions (since June 24, 2016).}. These containers were run on computers with one Intel i7-6700 SGX capable processor and 8 GB of RAM.

\subsection{The Experiments}
 
\begin{figure}[ht]
\centering
\includegraphics[width=.45 \textwidth]{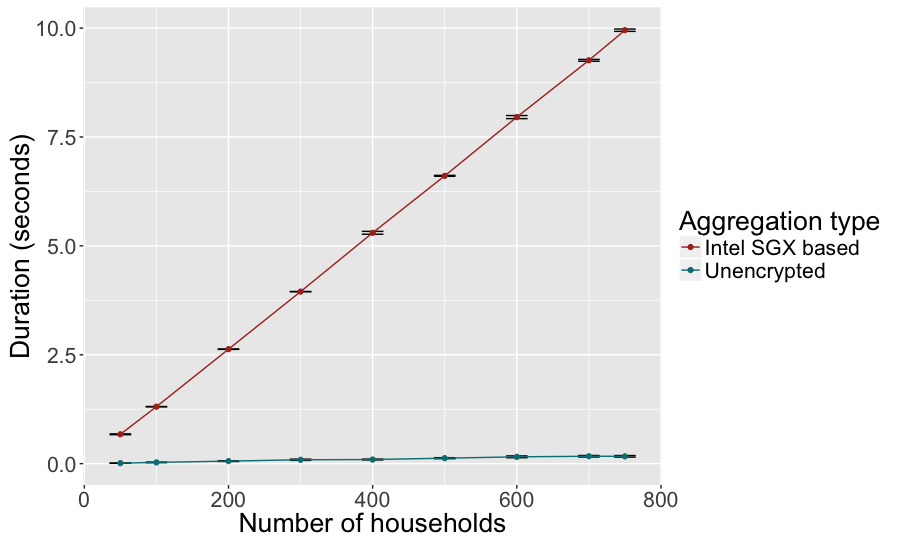}
\caption{Comparison between aggregating within SGX enclave and within regular code}
\label{fig:runtime} 
\end{figure}

The experiments consisted of two parts. The first part aimed to compare the execution time for the aggregation of daily consumption data for regions with $50$, $100$, $200$, $300$, $400$ and $500$, $600$, $700$ and $750$ households, sending measurements every 60 seconds. For every group size, $10$ runs were executed. Also, Figure~\ref{fig:runtime} shows the run times with a $95\%$   confidence interval. This experiment is important in order to find out the overhead caused by security measures.  

\begin{figure}[ht]
\centering
\includegraphics[width=.45 \textwidth]{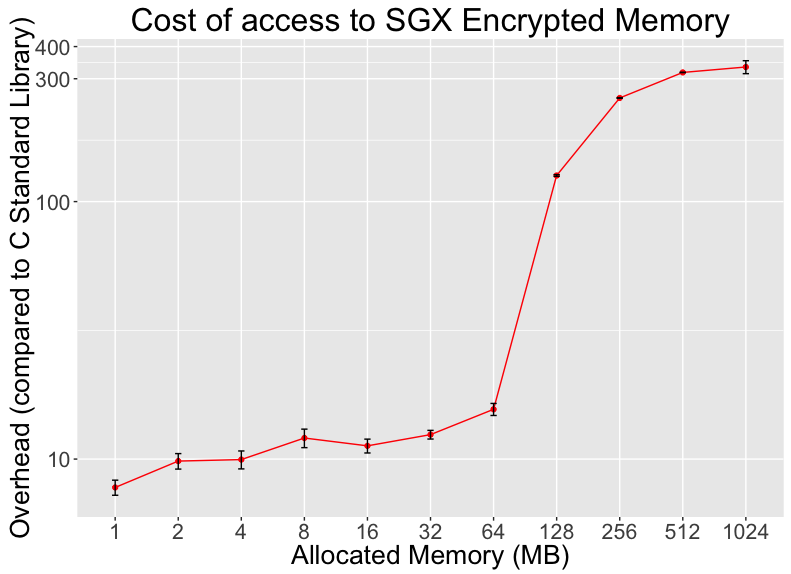}
\caption{Comparison between reading from SGX enclave and from unencrypted memory}
\label{fig:memory-overhead} 
\end{figure}

The second part had the objective of showing the relation between the memory access cost and the amount of memory allocated to an SGX enclave. The results can be seen in Figure~\ref{fig:memory-overhead} and show a considerable increase in access cost when the memory exceeds the size of the enclave page cache. This effect is a consequence of cache lines being paged out to main memory,  requiring (hardware-based) encryption as data leaves the processor.

\subsection{Discussion} 

As can be seen in Figure~\ref{fig:runtime}, the overhead of SGX is in the order of $40$ to $60$ times.
Despite the increased execution time, the benefits are significant because the privacy of the households are protected by the fact that the electricity consumption values are passed to the enclave within the aggregator along a secure channel established during the attestation process, and only the aggregated values will ever leave the enclave (as the code can be verified with the attestation).

On the other hand, it is important to keep in mind that the memory space that can be allocated to an enclave is currently limited to $128$ MB for all the enclaves within an Intel SGX platform, and, access times can increase significantly when this limit is reached, causing a different kind of pagination, where memory will be allocated outside the enclave in an encrypted way. This kind of pagination, as of September 2016, only works on Linux platforms. 

Nevertheless, for applications that require a small memory footprint as is our case, approximately $256$ bytes are needed to store accumulated data and cryptographic keys.
Therefore, one single enclave can be used to aggregate data from up to 500,000 smart meters without exceeding the enclave memory. Additional load distribution is also possible by balancing the SMs accross multiple SGX-capable machines.


\section{Conclusions} 
\label{sec:conclusions}

In this paper we have shown how Intel SGX technology can be used to provide a simple and general solution for the smart metering privacy problem. We provided a short introduction to a variety of previous solutions, as well as an account of the proposed Intel SGX based architecture providing a general and efficient solution to the problem. Finally, we presented also a set of experiments for assessing the performance of the proposed solution.

As future work we aim to assess other strategies for aggregation such as \textit{garbled circuits}~\cite{kolesnikov2008improved} and multi-party secure computing~\cite{cramer2000general}. 
Moreover, with regard to  Intel SGX we intend to integrate our implementation into cloud environments by, e.g., introducing mechanisms such as smooth remote attestation~\cite{barbosa2016foundations}. We also want to test how our architecture will perform using other secure hardware processing strategies, such as AMD Secure Memory Encryption (SME) and Secure Encrypted Virtualization (SEV).

\section*{Acknowledgements}

This research was partially funded by EU-BRA
SecureCloud project (MCTI/RNP 3rd Coordinated Call) and by CNPq, Brazil.

\bibliographystyle{abbrv}
\bibliography{sigproc}  
\end{document}